\documentstyle[prl,aps]{revtex}

\begin{document}
\title
{Soluble field theory with a massless gauge invariant limit}

\author {C. R. Hagen\cite{Hagen}}

\address
{Department of Physics and Astronomy\\
University of Rochester\\
Rochester, N.Y. 14627}

\maketitle

\begin{abstract}
     It is shown that there exists a soluble four parameter model in (1+1)
dimensions all of whose propagators can be determined in terms of the
corresponding known propagators of the vector coupling theory.  Unlike the
latter case, however, the limit of zero bare mass is nonsingular and yields a
nontrivial theory with a rigorously unbroken gauge invariance.
\end{abstract}

PACS number(s): 11.10Kk

Since quantum field theories in (3+1) dimensions have not been amenable to
exact calculation in any but the most trivial cases, the study of the
soluble field theories in (1+1) dimensions has been one of considerable
interest.  It originated over forty years ago with Thirring's observation [1]
that a current-current interaction should be exactly soluble in such a space
inasmuch as the current operator must have both vanishing divergence and curl.
A solution of the model was given by Johnson [2] who realized that an essential
ingredient had to be a very precise definition of the current operator.  He
adopted an approach in which the current is realized as an average of
spacelike and timelike limits of the product of two field operators.  Such an
averaging process allowed a covariant result to be obtained, but introduced the
somewhat undesirable feature of a timelike limit which does not fit comfortably
into a Cauchy initial value formulation.

The most general solution of the Thirring model was obtained by the author [3]
using an extension of Schwinger's gauge invariant definition of the current
$j^\mu$.  Specifically, one writes, in the case of a charged fermion $\psi (x)$
coupled to an external field $A_\mu(x)$[4]
\begin{eqnarray}
j^\mu(x) & =& \lim_{{\bf x}\to{\bf x'}}{1\over 2}\psi(x)q\alpha^\mu
\nonumber \\ &  &
exp[iq\int^x_{x'} dx^{\prime\prime}_\nu(\xi
A^\nu-\eta\gamma_5\epsilon^{\nu\alpha}A_\alpha)]\psi(x')
\end{eqnarray}
where the Dirac matrices $\alpha^0$ and $\alpha^1 = \gamma_5$
are conveniently taken to be
the unit matrix and the Pauli matrix $\sigma_3$, respectively, and the limit is
taken from a spacelike direction.  Although the parameters $\xi$ and $\eta$ are
required by Lorentz invariance to satisfy the constraint
$$\xi +\eta=1,$$
it is desirable for symmetry reasons to retain both parameters in the general
formulation.  Worth noting is that $\xi=1$($\eta=1$) corresponds to vector
(axial-vector) conservation while the choice $\xi=\eta={1\over 2}$ turns out to
be equivalent to the Johnson solution.

Among the most well known of the two-dimensional models is the Schwinger model
[5] which is simply QED for a massless fermion.  Its solution finally put to
rest the widely held view that gauge invariance necessarily required a
massless photon.  The extension of the Schwinger model to the massive vector
meson case was carried out by Sommerfield [6], Brown [7], and the author [8]
for the cases $\xi={1\over 2}$, $\xi=1$, and arbitrary $\xi$ respectively.

A model in which only a single component of the fermion field was coupled via
the current operator to a massive vector meson was subsequently proposed and
solved by the author [9].  This single component model has the same Green's
functions as one which was subsequently proposed by Jackiw and Rajaraman [10].
The latter formulation has been referred to as the chiral Schwinger model, a
somewhat unfortunate description inasmuch as the propagator equations of
motion are inconsistent with zero bare mass for the photon [11].

The final extension of the model to date was accomplished with the proposal
[12] that the model be generalized to include arbitrary admixture of vector
and axial vector couplings.  This was subsequently solved [13] with the
definition (1) generalized to
\begin{eqnarray}
\lefteqn{j^\mu = \lim_{{\bf x}\to{\bf x'}}
{1\over 2}\psi(x)q\alpha^\mu (1+r\gamma_5)} \nonumber \\
& & exp [iq\int^x_{x'}
dx^{\prime\prime}_\nu(\xi A^\nu-\eta\gamma_5\epsilon^{\nu\alpha}A_\alpha
\nonumber \\
& & +r\eta\gamma_5A^\nu-r\xi\epsilon^{\nu\alpha}A_\alpha)]\psi(x')
\end{eqnarray}
where $er$ is the axial-vector coupling constant.
Using (2) in conjunction with the Lagrangian
\begin{eqnarray}
\cal L  & = &{i\over 2}\psi\alpha^\mu\partial_\mu\psi+{1\over 4}G^{\mu\nu}
G_{\mu\nu}-{1\over 2}G^{\mu\nu}(\partial_\mu B_\nu-\partial_\nu B_\mu)
\nonumber \\
& \quad - &{1\over 2}
\mu_0^2B^\mu B_\mu+ej^\mu B_\mu+j^\mu A_\mu+J^\mu B_\mu,
\end{eqnarray}
functional derivatives with respect to the external sources $A_\mu$ and $J_\mu$
can be used to obtain a complete solution.  Worth noting is the fact that this
model has four parameters two of which (namely, $e$ and $\mu_0$) appear
explicitly in the Lagrangian while two more ($\xi$ and $r$) occur only in the
definition of the current operator (2).

Perhaps the most striking feature of the solution is the mass spectrum of the
model.  In addition to a massless mode there is a renormalization of the mass
$\mu_0$ which is given by
\begin{equation}
\mu^2=\mu_0^2{[1+(\xi e^2/\pi\mu_0^2)(1-r^2)][1-(\eta e^2/\pi\mu_0^2)(1-r^2)]
\over 1-(e^2/\pi\mu_0^2)(\xi r^2+\eta),}
\end{equation}
in agreement with results obtained in refs.[1-3,5-11] for special cases of the
four parameters of the model.  Of particular interest is the case $r^2=1$
considered in [9] and [10] for which the physical mass reduces to
$$\mu^2=\mu_0^2\bigg[1-{e^2\over \pi\mu_0^2}\bigg]^{-1}.$$
This result clearly precludes one of the most interesting possible limits,
namely the case $\mu_0=0$ (i.e., a chiral theory with gauge invariance).
While the very different calculational approach of [10] argues for the
possibility of that limit, it concludes with the essentially equivalent result
that gauge invariance is necessarily broken as a consequence of an anomaly.
In this work a new model is proposed which, like that of [13], is a four
parameter one, but which allows the chiral $r^2=1$, $\mu_0=0$ limit without
either destroying gauge invariance or requiring tachyonic modes.

One begins with the Lagrangian
\begin{eqnarray}
\lefteqn{{\cal L}_S  = {i\over 2}\psi\alpha^\mu\partial_\mu\psi+
{1\over 2}\varphi^\mu\varphi_\mu +\varphi^\mu\partial_\mu\varphi}\nonumber \\
& &-{1\over 2}\mu_0^2\varphi^2
+gj^\mu\varphi_\mu+j^\mu{\cal A}_\mu+{\cal J}^\mu\varphi_\mu
\end{eqnarray}
which for $g=0$ and vanishing sources ${\cal A}_\mu$ and ${\cal J}_\mu$ is
recognized to be the Lagrangian of a massless fermion coupled to a scalar field
of mass $\mu_0$.  The equations of motion have the form
$$\alpha^\mu({1\over i}\partial_\mu-gq\varphi_\mu-q{\cal A}_\mu)\psi=0$$
$$\varphi_\mu=-\partial_\mu\varphi-gj_\mu-{\cal J}_\mu$$
$$\partial_\mu\varphi^\mu+\mu_0^2\varphi=0$$
while the nonvanishing equal time commutators are
$$[\varphi^0(x),\varphi(x')]=-i\delta ({\bf x}-{\bf x'})$$
and
$$\{\psi (x),\psi (x')\}=\delta ({\bf x}-{\bf x'}).$$
To display the close connnection between (3) and (5) the former is rewritten as
${\cal L}_V$ which is then expressed in terms of new fields $\phi$ and
$\phi^\mu$ according to
$$G^{\mu\nu}=\mu_0\epsilon^{\mu\nu}\phi$$
$$B^\mu={1\over \mu_0}\epsilon^{\mu\nu}\phi_\nu$$
thereby yielding
\begin{eqnarray*}
\lefteqn{{\cal L}_V = {i\over 2}\psi\alpha^\mu\partial_\mu\psi+
{1\over 2}\phi^\mu\phi_\mu+\phi^\mu\partial_\mu\phi}  \\
& &-{1\over 2}\mu_0^2\phi^2
+{e\over \mu_0}j_\mu\epsilon^{\mu\nu}\phi_\nu+j^\mu A_\mu+{1\over \mu_0}
\epsilon^{\mu\nu}J_\mu\phi_\nu.
\end{eqnarray*}

Upon using the identity
$$\epsilon^{\mu\nu}\alpha_\nu(1+r\gamma_5)=r\alpha^\mu(1+{1\over r}\gamma_5)$$
and the fact that the exponential which appears in the definition (2) of the
current operator is invariant under the simultaneous replacements $r\to r^{-1}$
and $A^\mu\to-r\epsilon^{\mu\nu}A_\nu$, one infers that
$$\epsilon^{\mu\nu}j_\nu(r,A^\mu)=rj^\mu({1\over r},-r\epsilon^{\alpha\beta}
A_\beta).$$
This allows ${\cal L}_V$ to be rewritten as
\begin{eqnarray*}
\lefteqn{{\cal L}_V={i\over 2}\psi\alpha^\mu\partial_\mu\psi+
{1\over 2}\phi^\mu\phi_\mu+\phi^\mu\partial_\mu\phi
-{1\over 2}\mu_0^2\phi^2 } \\
 & & -{er\over \mu_0}j^\mu({1\over r},
-r\epsilon^{\alpha\beta}A_\beta)\phi_\mu-
j^\mu({1\over r},-r\epsilon^{\alpha\beta}A_\beta)r\epsilon_{\mu\nu}A^\nu \\
& \quad & +{1\over \mu_0}\epsilon^{\mu\nu}J_\mu\phi_\nu
\end{eqnarray*}
thereby establishing a complete equivalence between the two models.
Specifically, the vacuum-to-vacuum transition amplitudes $\langle 0\sigma_1
|0 \sigma_2 \rangle^{S,V}$ are seen to satisfy the relationship
\begin{eqnarray}
\langle 0 \sigma_1 & | & 0 \sigma_2 \rangle^{S,g,r}({\cal A}^\mu,{\cal
J}^\mu)  = \nonumber  \\
& & \langle 0 \sigma_1|0 \sigma_2\rangle^{V,e=-gr\mu_0,{1\over r}}
(-r\epsilon^{\mu\nu}{\cal A}_\nu,-\mu_0\epsilon^{\mu\nu}{\cal J}_\nu).
\end{eqnarray}
Corresponding results for the $2n$-point fermion propagators $G^{S,g,r}
(x_1,...x_{2n};{\cal A}^\mu,{\cal J}^\mu)$ can be written in the form
\begin{eqnarray}
&G&^{S,g,r}(x_1,...x_{2n};{\cal A}^\mu,{\cal J}^\mu) \nonumber \\
& & =  G^{V,e=-gr\mu_0,{1\over
r}}(x_1,...x_{2n};-r\epsilon^{\mu\nu}{\cal A}_\nu,-\mu_0\epsilon^{\mu\nu}{\cal
J}_\nu).
\end{eqnarray}
Clearly Eq.(6) allows a calculation of the nonvanishing bosonic mass in the
model by letting $r\to {1\over r}$ and $e\to -{gr\mu_0}$ in (4).
The result is
\begin{equation}
\mu^2=\mu_0^2{[1-(\xi g^2/\pi)(1-r^2)][1+(\eta g^2/\pi)(1-r^2)]\over
1-(g^2/\pi)(\xi +\eta r^2)},
\end{equation}
thereby clearly displaying the nonsingularity of the $\mu_0\to 0$ limit [14].
It should be noted that all of the usual limits which can be taken in the model
described by ${\cal L}_V$ have their counterparts in the ${\cal L}_S$ theory.
Thus, for example, a three parameter Thirring model is obtainable in the
$\mu_0\to\infty$ limit with the more usual two parameter model emerging in the
subsequent limits $r=0$ and $r\to\infty, gr$ finite.

To illustrate how the calculation of matrix elements is to be carried out one
can simply note that the current correlation function $D_S^{\mu\nu}(x-x')$
for example is given by
$$D_S^{\mu\nu}(x-x')=-i{\delta^2\over \delta {\cal A}_\mu(x)\delta {\cal A}_
\nu(x')}ln\langle 0 \sigma_1|0 \sigma_2\rangle^S|_{{\cal A}={\cal J}=0}$$
which by (6) reduces to
$$D^{\mu\nu}_S(p)=r^2\epsilon^{\mu\alpha}\epsilon^{\nu\beta}
D^V_{\alpha\beta}(p)|^{1/r}_{e=-gr\mu_0}.$$
This readily yields from [13] the result
\begin{eqnarray*}
D^{\mu\nu}_S(p) & = & (g^{\mu\alpha}+r\epsilon^{\mu\alpha})
(g^{\nu\beta} +r\epsilon^{\nu\beta})
\left[ D_1(p)\epsilon_{\alpha\sigma}
p^\sigma \epsilon_{\beta\tau}p^\tau \right.\\
& & \left. + D_2(p)p_\alpha p_\beta+D_3(p)(p_\alpha\epsilon_{\beta\sigma}
p^\sigma+p_\beta \epsilon_{\alpha\sigma}p^\sigma) \right]
\end{eqnarray*}
where
\begin{eqnarray*}
D_1(p)& = &{\xi\over \pi}{1\over 1-{\xi g^2\over \pi}(1-r^2)}
 \biggl[  D(p)+ \\
 & & \rule{0in}{.30in}
  {g^2\xi r^2\over \pi}
{1+{\eta g^2\over \pi}(1-r^2)\over 1-{g^2\over \pi}(\xi+\eta r^2)}
\Delta (p) \biggr]
\end{eqnarray*}
\begin{eqnarray*}
D_2(p)& = & {\eta\over \pi}{1\over 1+{\eta g^2\over \pi}(1-r^2)}
\biggl[  D(p)+ \\
& & \rule{0in}{.30in}
 {g^2\eta\over \pi}
{1-{g^2\xi\over \pi}(1-r^2)\over 1-{g^2\over \pi}(\xi+\eta r^2)}
\Delta(p) \biggr]
\end{eqnarray*}
$$D_3(p)=-{g^2\xi\eta r\over \pi}{1\over 1-{g^2\over \pi}(\xi+\eta r^2)}
\Delta(p)$$
with $D(p)=1/p^2$, $\Delta(p) = 1/(p^2+\mu^2)$, and $\mu^2$ given by Eq.(8).

Also of considerable interest are the propagators of the fields $\varphi^\mu$
and $\varphi$.  The former is obtained from the corresponding result of the
vector theory[13] according to
$$G_S^{\mu\nu}(p)=\mu_0^2\epsilon^{\mu\alpha}\epsilon^{\nu\beta}G_{\alpha\beta}
^V(p)|^{1/r}_{e=-gr\mu_0}.$$
This yields
\begin{eqnarray*}
\lefteqn{ G_S^{\mu\nu}(p)  =
{\mu_0^2\over \mu^2}[p^\mu p^\nu\Delta(p)-g^{\mu\nu}] } \\
 & & \rule{0in}{.30in}
\quad + {g^2\over \pi}{1\over [1-{\xi g^2\over \pi}(1-r^2)]
[1+ {\eta g^2\over\pi}(1-r^2)]} \\
 & & \rule{0in}{.30in}
 \quad \biggl\{ \frac{ } { } [-r(p^\mu\epsilon^{\mu\alpha}p_\alpha +
 p^\nu\epsilon^{\mu\alpha}
p_\alpha)+(1+r^2)\epsilon^{\mu\alpha}p_\alpha\epsilon^{\nu\beta}p_\beta]
  \\
 & & \rule{0in}{.30in} \quad
 (D-\Delta) + \epsilon^{\mu\alpha}p_\alpha\epsilon^{\nu\beta}p_\beta{1+r^2-
{g^2\over\pi}(\xi + \eta r^4)\over 1-{g^2\over \pi}(\xi +\eta r^2)} \Delta
\biggr\}.
\end{eqnarray*}
Upon using the equations of motion and the commutation relations [15] in the
case $\mu_0\neq 0$ the $\varphi$ propagator is then found to be
$$G_S(p)={1\over \mu_0^2}+p_\mu p_\nu G_S^{\mu\nu}(p),$$
which readily reduces to
$$G_S(p)={\mu^2\over \mu_0^2}\Delta(p).$$
Thus the $\varphi$ propagator is seen to differ from that of the free theory
merely by wave function and mass renormalizations.

Since the propagators of the model are all finite in the limit of vanishing
bare mass, the solution of the $\mu =0$ case can be obtained merely by taking
this limit.  In some respects this is the analogue of the Schwinger model but
differs in that by Eq.(8) the physical mass is zero.  A further point of
difference is that in the vector theory there is consistency only for the cases
$r=\eta =0$ (the usual Schwinger model) and $r\to\infty, er$ finite, $\xi=0$
(the axial Schwinger model).  In the theory presented here no restrictions need
to be placed on the parameters $g,r,$ and $\xi$ and the propagators reduce to
the form
$$G_S^{\mu\nu}(p)={1\over 1-{g^2\over \pi}(\xi +\eta r^2)}(p^\mu p^\nu
-p^2g^{\mu\nu})D(p)$$
and
$$G_S(p)={[1-(\xi g^2/\pi)(1-r^2)][1+(\eta g^2/\pi)(1-r^2)]\over 1-(g^2/\pi
)(\xi +\eta r^2)}D(p).$$
This is clearly a massless gauge theory which has the property of being
invariant under the gauge transformation
$$\varphi\to\varphi +\Phi_0$$
where $\Phi_0$ is a constant[16].  The theory thus supports the broken
symmetry condition[17]
$$\langle 0|\varphi (x)|0 \rangle =\Phi _0,$$
in stark contrast to the model considered in [9] and [10].  It should also be
noted that there is no difficulty in passing to the chiral limit $r^2=1$ with
the particularly simple results
$$G_S^{\mu\nu}(p)={1\over 1-(g^2/\pi)}(p^\mu p^\nu -p^2g^{\mu\nu})D(p)$$
and
$$G_S(p)={1\over 1-(g^2/\pi )}D(p).$$
The fermionic two point function for this massless chiral theory is readily
seen from Eq.(7) to be
\begin{eqnarray*}
\lefteqn{ G^{S,g,r}(x-x';{\cal A}={\cal J}=0) } \\
 & =& exp\{-ig^2{1\over1-(g^2/
 \pi)}
 (1\pm\gamma_5)^2[D(x-x')-D(0)]\}.
\end{eqnarray*}

In conclusion it should be remarked that the model presented here has the
advantage of being fully soluble and contains a wealth of submodels obtainable
by taking appropriate limits in the relevant parameter space. In contrast to
the case of the related vector models where the limit of vanishing bare mass
and chiral currents leads to inconsistencies and/or breaking of the gauge
symmetry, no such problems are encountered in the theory presented here.  This
leads one to hope that it may well prove to be a more reliable theoretical
laboratory than some of the alternative soluble two dimensional theories.

\acknowledgments

This work is supported in part by the U.S. Department of Energy Grant
No.DE-FG02-91ER40685.


\begin{references}
\bibitem{ } In order to simplify the results for the propagators one uses here
a charge matrix
$$q=\pmatrix{0&-i\cr
i&0\cr}$$
which acts in a two-dimensional charge space of the Hermitian field $\psi(x)$.
\bibitem{ } C. R. Hagen, Phys. Rev. D{\bf 55}, 1021 (1997).  It should be
remarked here that the result quoted there for $M^{\mu\nu}(p)$ needs to be
multiplied by a factor of $\mu_0^2$.
\bibitem{ } Equation (6) suggests that the dependence upon $\mu_0$ is
essentially trivial, serving only to set the mass scale.  The complete
solution of the model in fact confirms this observation (i.e., $\mu_0$ serves
no purpose in the model other than as a scale factor for the mass).
\bibitem{ } The latter is required for the evaluation of a term which arises
from the derivative of the time ordered product.
\end{references}
\end{document}